\documentclass[a4paper,11pt]{article}
\usepackage{pos}

\usepackage{amsmath}
\usepackage{graphicx}
\usepackage{multirow}
\usepackage{slashed}

\newcommand{\be}{\begin{equation}} \newcommand{\ee}{\end{equation}}
\newcommand{\ba}{\begin{array}{c}} \newcommand{\ea}{\end{array}}
\newcommand{\bea}{\begin{eqnarray}} \newcommand{\eea}{\end{eqnarray}}

\allowdisplaybreaks[4]

\title{A lattice QCD study of low-energy interactions of doubly charmed baryons}
\author*[a,b]{Jing-Yu~Yi}
\author[c]{Ze-Rui~Liang}
\author[d,e]{Liuming~Liu}
\author[a,b]{De-Liang~Yao}

\affiliation[a]{School of Physics and Electronics, Hunan University, 
Changsha 410082, China}
\affiliation[b]{Hunan Provincial Key Laboratory of High-Energy Scale Physics and Applications,\\ Hunan University,
410082 Changsha, China}
\affiliation[c]{College of Physics and Hebei Key Laboratory of Photophysics Research and Application, \\ Hebei Normal University, 
Shijiazhuang 050024, Hebei, China}
\affiliation[d]{Institute of Modern Physics, Chinese Academy of Sciences,
 Lanzhou 730000, China}
\affiliation[e]{University of Chinese Academy of Sciences,
Beijing 100049, China}

\emailAdd{yijingyu@hnu.edu.cn}
\emailAdd{liangzr@hebtu.edu.cn}
\emailAdd{liuming@impcas.ac.cn}
\emailAdd{yaodeliang@hnu.edu.cn}

\abstract{We present a lattice QCD calculation of the S-wave interactions between the spin-1/2 doubly charmed baryons ($\Xi_{cc},\Omega_{cc}$) and Goldstone bosons ($\pi,K,\bar{K}$) using the $N_f=2+1$ CLQCD ensembles with lattice spacing $a = 0.07746$~fm and two pion masses $M_\pi \simeq 210$ and 300 MeV. Four single-channel interactions, $I=1/2$ $\Omega_{cc}\bar{K}$, $I=1$ $\Xi_{cc}K$, $I=0$ $\Xi_{cc}K$ and $I=3/2$ $\Xi_{cc}\pi$, are investigated, since they are free of disconnected diagrams. Lüscher's finite volume method is utilized to extract the scattering parameters in the effective range expansion, i.e., the scattering lengths and effective ranges, from the finite-volume spectra. Our obtained results agree well with previous baryon chiral perturbation theory predictions.}

\FullConference{The 21st International Conference on Hadron Spectroscopy and Structure (HADRON2025)\\
 27 - 31 March, 2025\\
Osaka University, Japan\\}

\begin{document}
\maketitle

\section{Introduction}
The study of doubly charmed baryons (DCBs), $\Xi_{cc}^{++}$, $\Xi_{cc}^{+}$, and $\Omega_{cc}^{+}$, plays a crucial role in our understanding of low-energy dynamics of quantum chromodynamics (QCD) in the heavy-quark sector. The DCBs comprise two charm quarks and one light quark, and hence provide a unique ground to explore both heavy-quark and chiral symmetry breakings. On the experimental side, the discovery of $\Xi_{cc}^{++}$ with a mass near 3621 MeV by LHCb marks a major milestone~\cite{LHCb:2017iph}, followed by measurements of its lifetime~\cite{LHCb:2018zpl} and production cross-section~\cite{LHCb:2019qed}. Though various decay modes of $\Xi_{cc}^{++}$ are found and well measured~\cite{LHCb:2017iph, LHCb:2018pcs, LHCb:2022rpd, LHCb:2025shu}, the $\Xi_{cc}^{+}$, $\Omega_{cc}^{+}$, and DCB excited states remain unobserved,  limiting our understanding of DCB properties and interactions.

On the theoretical aspect, a plenty of works have been devoted to the exploration of DCBs ~\cite{Lu:2017meb, Wang:2017mqp, Chen:2017sbg, Wang:2018lhz, Shi:2019hbf, Cheng:2020wmk, PACS-CS:2013vie, Alexandrou:2017xwd, Can:2019wts, Bahtiyar:2020uuj, Alexandrou:2023dlu}, though the lack of experimental data poses a big challenge to theorists. So far, there are no experimental and lattice QCD data for the interactions between DCBs and Goldstones (GBs). The GBs are also denoted by $\phi\in\{\pi,K,\bar{K},\eta\}$ hereafter. The absence of scattering data forces investigations within baryon chiral perturbation theory (BChPT) to rely on approximations, such as heavy diquark-antiquark (HDA) symmetry, to estimate the values of the involved low-energy constants (LECs). While lattice QCD has achieved precise determinations of hadron spectroscopy~\cite{PACS-CS:2013vie, Alexandrou:2017xwd, Can:2019wts, Bahtiyar:2020uuj, Alexandrou:2023dlu}, first-principle calculations of DCB-$\phi$ scatterings remain absent.

In this proceeding, we present a lattice QCD analysis of S-wave scattering for the four channels, denoted by (strangeness, isospin) as $\Xi_{cc}\pi(0,3/2)$, $\Xi_{cc}K(1,0)$, $\Xi_{cc}K(1,1)$, and $\Omega_{cc}\bar{K}(-2,1/2)$, as an update of the lattice results shown in Ref.~\cite{Liang:2025kkn}. With Lüscher's method~\cite{Luscher:1990ux}, we determine scattering lengths and effective ranges in the effective range expansion (ERE)~\cite{Bethe:1949yr, Blatt:1949zz}. Our study provides crucial lattice QCD data for future studies of double-heavy spectroscopy and for a direct test of the HDA symmetry assumptions used in BChPT analyses of the DCBs.

\section{lattice setup and finite-volume spectrum}
We employ the same gauge configurations, generated by the CLQCD collaboration~\cite{CLQCD:2023sdb}, as in our previous study~\cite{Liang:2025kkn}. There are four ensembles with the same lattice spacing $a = 0.07746$~fm and two different pion masses $M_{\pi}\simeq 210$ MeV and $M_{\pi}\simeq 300$ MeV; see Ref.~\cite{Liang:2025kkn} for details of the ensembles. For the valence charm quark, we implement the Fermilab action~\cite{El-Khadra:1996wdx}, with parameters tuned according to the procedure established in Ref.~\cite{Liu:2009jc}. The quark propagators are computed with the distillation quark smearing method~\cite{HadronSpectrum:2009krc}. The number of eigenvectors is 100 for all four ensembles.  
The interpolating operators for GBs and DCBs are defined in Ref.~\cite{Liang:2025kkn}. Explicitly, the single-particle operators are constructed as follows:
\begin{align}
\mathcal{O}_{\pi^+}(x)&=\bar{d}(x)^a_{\alpha} (\gamma_5)_{\alpha \beta} u(x)^a_{\beta}, 
\quad 
\mathcal{O}_{\Xi_{cc}^{++}(ccu)}(x)=\epsilon^{ijk}P_{+}[Q_c^{i T}(x)C\gamma_5 q_u^j(x)] Q_c^k(x),  \\
\mathcal{O}_{K^+}(x)&=\bar{s}(x)^a_{\alpha } (\gamma_5)_{\alpha \beta} u(x)^a_{\beta},
\quad 
\mathcal{O}_{\Xi_{cc}^+(ccd)}(x)=\epsilon^{ijk}
P_{+}[Q_c^{i T}(x) C\gamma_5 q_d^j(x)] Q_c^k(x),
 \\
\mathcal{O}_{K^0}(x)&=\bar{s}(x)^a_{\alpha} (\gamma_5)_{\alpha \beta} d(x)^a_{\beta},\quad
\mathcal{O}_{\Omega_{cc}^{+}(ccs)}(x)=\epsilon^{ijk}
P_{+}[Q_c^{i T}(x) C\gamma_5 q_s^j(x)] Q_c^k(x),
\end{align}
where $C$ is the charge conjugation matrix and $P_+=(1+\gamma_0)/2$ is the parity projector. For two-particle systems, we construct the operators in the rest frame as
$\mathcal{O}(\mathbf{p}) = \sum_{\alpha, \mathbf{p}} c_{\alpha, \mathbf{p}} B_{\alpha}(\mathbf{p}) M(-\mathbf{p})$, where $B$ and $M$ represent the baryon and meson single-particle operators, respectively. The operators with definite isospins for the four channels read
\begin{align}
\mathcal{O}_{\mathbf{p}}
^{\Xi_{c c} K,\ I=0}
&=\sum_{\alpha, \mathbf{p}} C_{\alpha, \mathbf{p}}
\left(
\frac{1}{\sqrt{2}} \mathcal{O}_{\Xi_{c c}^{++}, \alpha}(\mathbf{p}) \mathcal{O}_{K^0}(-\mathbf{p})-\frac{1}{\sqrt{2}} \mathcal{O}_{\Xi_{c c}^{+}, \alpha}(\mathbf{p}) \mathcal{O}_{K^{+}}(-\mathbf{p})
\right),\\
\mathcal{O}_{\mathbf{p}}
^{\Xi_{c c} K,\ I=1}
&= \sum_{\alpha, \mathbf{p}} C_{\alpha, \mathbf{p}}
\left(
\mathcal{O}_{\Xi_{c c}^{++}, \alpha}(\mathbf{p}) \mathcal{O}_{K^+}(-\mathbf{p})
\right),\\
\mathcal{O}_{\mathbf{p}}
^{\Omega_{c c} \bar{K},\ I=1/2}
&=\sum_{\alpha, \mathbf{p}} C_{\alpha, \mathbf{p}}
\left(
\mathcal{O}_{\Omega_{c c}^{+}, \alpha}(\mathbf{p}) \mathcal{O}_{\bar{K}^0}(-\mathbf{p})
\right),
\\
\mathcal{O}_{\mathbf{p}}
^{\Xi_{c c} \pi,\ I=3/2}
&= \sum_{\alpha, \mathbf{p}} C_{\alpha, \mathbf{p}}
\left(
\mathcal{O}_{\Xi_{c c}^{++}, \alpha}(\mathbf{p}) \mathcal{O}_{\pi^+}(-\mathbf{p})
\right).
\end{align}
Here, $\alpha$ denotes the Dirac index of the baryon operator and $\bf{p}$ is the momentum. The coefficients $C_{\alpha, \bf{p}}$ are chosen such that the operator transforms in the $G_1^-$ irreducible representation of the cubic group, corresponding to the quantum numbers $J^P = \frac{1}{2}^-$ of the scattering channels under study. For each channel, we employ three operators with $|\mathbf{p}| = 0, 1, \sqrt{2}$ (in units of $2\pi/L$). Explicit values of the coefficients $C_{\alpha, \bf{p}}$ can be found in Ref.~\cite{Xing:2022ijm}.

Single-particle masses are determined from fits to their respective correlation functions, with results summarized in Table~\ref{tab:single.particle.meff}. The dispersion relations are examined through fits to $E^2=m_0^2 +c^2\mathbf{p}^2$ for the five lowest lattice momenta. The results are summarized in Table~\ref{tab:single.particle.meff.DR}. For most particles, the results exhibit good agreement with relativistic dispersion, though $\Xi_{cc}$ and $\Omega_{cc}$ display minor deviations in specific ensembles due to heavy-quark discretization effects. For two-particle systems, we compute $3\times3$ correlation matrices and extract finite-volume energy levels through the generalized eigenvalue problem (GEVP)~\cite{Luscher:1990ck}. The GEVP analysis employs a reference timeslice $t_0=4$ in lattice units, and the resulting eigenvalues are fitted to extract the energy spectrum. The obtained energy levels reveal distinctive interaction patterns across different channels.

\begin{table}[hb]
\begin{center}
\caption{Masses of single particles extracted from fits to correlation functions, presented in GeV. Results are shown for all the four ensembles.}
\label{tab:single.particle.meff}
\begin{tabular}{c|c|c|c|c}
\hline
\hline
      &     $\pi$&       $K$& $\Xi_{cc}$& $\Omega_{cc}$\\
\hline    
F32P30&  0.3039(6)& 0.5230(4)&  3.6324(9)&     3.7137(6)
\\
\hline 
F48P30&  0.3049(4)& 0.5241(3)& 3.6372(13)&    3.7179(10)
\\
\hline 
F32P21& 0.2081(19)& 0.4917(7)& 3.6056(14)&     3.6938(8)
\\
\hline 
F48P21&  0.2074(7)& 0.4911(3)& 3.6076(18)&    3.6997(10)
\\
\hline
\hline
\end{tabular}
\end{center}
\end{table}

\begin{table}[ht]
\begin{center}
\caption{Parameters from fits to the dispersion relation $E^2 = m_0^2 + c^2\mathbf{p}^2$ for single particles. The rest mass $m_0$ and the coefficient $c^2$ are given for each particle across the four ensembles.\label{tab:single.particle.meff.DR}}
\renewcommand{\arraystretch}{1.1}
\begin{tabular}{l|c|r|r|r|r}
\hline\hline
&& F32P30 & F48P30 & F32P21 & F48P21 \\
\hline
{$\pi$} & $m_0~\text{[GeV]}$ & $0.3040(6)$  & $0.3049(4)$ & $0.2085(19)$  & $0.2077(7)$ \\
& $c^2$ & $1.0069(42)$ & $1.0075(22)$ & $1.0582(94)$ & $1.0157(37)$\\
\hline
$K$&$m_0~\text{[GeV]}$&$0.5230(4)$&$0.5241(3)$&$0.4917(7)$&$0.4911(3)$\\
&$c^2$&$1.0026(23)$&$1.0034(15)$&$1.0097(52)$&$1.0081(25)$\\
\hline
$\Xi_{cc}$&$m_0~\text{[GeV]}$&$3.6330(8)$&$3.6369(12)$&$3.6055(14)$&$3.6080(16)$\\
&$c^2$&$0.9915(61)$&$0.9641(186)$&$1.0056(101)$&$1.0520(218)$\\
\hline
$\Omega_{cc}$&$m_0~\text{[GeV]}$&$3.7139(6)$&$3.7179(10)$& $3.6938(8)$&$3.6995(10)$\\
&$c^2$&$0.9747(37)$&$0.9665(205)$&$0.9801(51)$&$0.9976(131)$\\
\hline\hline
\end{tabular}
\end{center}
\end{table}

At both pion masses, the $\Xi_{cc}K(1,0)$ channel exhibits negative energy shifts relative to the non-interacting threshold, indicating attractive interactions. In contrast, the $\Xi_{cc}\pi(0,3/2)$, $\Xi_{cc}K(1,1)$, and $\Omega_{cc}\bar{K}(-2,1/2)$ channels show positive energy shifts, characteristic of repulsive dynamics. We exclude high-lying states near open channels from subsequent scattering analysis, due to potential coupling effects that could complicate the interpretation.

\subsection{Scattering analysis} 
We extract the infinite-volume S-wave scattering parameters in the ERE using Lüscher's finite-volume method. For the single channel S-wave scattering, the L\"uscher's formula associates the finite-volume energy with the infinite volume scattering phase shift $\delta_0$ via
\be
p\cot \delta_0(p) = \frac{2}{L\sqrt{\pi}} \mathcal{Z}_{00}(1; q^2)\ ,
\label{Eq:Luscher}
\ee
where $p$ is the scattering momentum obtained from the two-particle finite-volume energy $E = \sqrt{m_1^2 + p^2} +  \sqrt{m_2^2 + p^2}$, and $q = {pL}/{(2\pi)}$. The zeta-function $\mathcal{Z}_{00}(1; q^2)$ can be evaluated numerically provided $q^2$ is given. 
 The scattering length $a_0$ and effective range $r_0$ are then obtained by fitting $p\cot{\delta_0(p)}$ with the following ERE formula
\begin{equation}
p \cot\delta_0(p) = \frac{1}{a_0} + \frac{1}{2} r_0 p^2.
\label{eq:ERE_form}
\end{equation}

To address discretization artifacts from the heavy charm quarks, we apply a dispersion-relation correction (DRC) to the two-particle energies:
\begin{align}
     E_p^{\prime}=E_p+\left(M_{1(p)}^{\rm cont.}+M_{2(p)}^{\rm cont.}\right)-\left(M_{1(p)}^{\rm lat.}+M_{2(p)}^{\rm lat.}\right) \ . 
\end{align}
Here $E_{p}$ is the two-particle energy that has a dominant contribution from the operator $B(\mathbf{p})M(\mathbf{p})$. Furthermore, $M_{1(p)}^{\mathrm{cont.}}$ and $M_{2(p)}^{\mathrm{cont.}}$ are the energies of the two particles at momentum $p$ calculated from the continuum dispersion relation, respectively; $M_{1(p)}^{\mathrm{lat.}}$ and $M_{2(p)}^{\mathrm{lat.}}$ are the corresponding energies computed on the lattice. Our analysis shows that the scattering length $a_0$ is robust against this correction. This is because $a_0$ is primarily constrained by near-threshold energy levels, where the correction term vanishes. In contrast, the effective range $r_0$ exhibits significant sensitivity to this correction across most channels. 

It is noteworthy that the statistical uncertainties in this analysis are estimated using the bootstrap resampling method. This represents a refinement in the error analysis compared to the previous preliminary study~\cite{Liang:2025kkn}, leading to slightly updated central values and uncertainties for the scattering lengths. The S-wave scattering lengths $a_0$ for the four channels at $M_\pi \simeq 210$ MeV and 300 MeV are summarized in Table~\ref{tab:ScatLen}. Predictions from BChPT using the EOMS and heavy baryon (HB) schemes at the physical pion mass are shown for easy comparison. Our lattice QCD determinations are in good agreement with the BChPT predictions within uncertainties, even though they are obtained at the unphysical pion masses.

\begin{table}[ht] 
\begin{center}
\caption{S-wave scattering lengths $a_0$ for the four channels at $M_\pi \simeq 210$ MeV and 300 MeV, in units of fm. For comparison, the last two columns show BChPT predictions using the EOMS renormalization scheme~\cite{Liang:2023scp} and HB formalism~\cite{Meng:2018zbl} at the physical pion mass.}
\label{tab:ScatLen}
\renewcommand{\tabcolsep}{0.6pc}
\renewcommand{\arraystretch}{1.2}
\begin{tabular}{cc|cc|cc}
\hline\hline
$(S,I)$&Processes& $M_{\pi}\sim300$ MeV& $M_{\pi}\sim210$ MeV & EOMS&HB \\
\hline 
$(-2,\frac{1}{2})$&$\Omega_{cc} \bar{K} \to \Omega_{cc} \bar{K}$&$-0.162(20)$&$-0.136(12)$&$-0.09_{-0.13}^{+0.12}$ & -0.20(1) \\
\hline
$(1,1)$&$\Xi_{cc} K \to \Xi_{cc} K$ &$-0.177(22)$&$-0.212(14)$&$-0.60 \pm 0.13$ &$-0.25(1)$ \\
\hline
$(1,0)$&$\Xi_{cc} K \to \Xi_{cc} K$ &$0.63(10)$&$0.697(90)$&$1.03 \pm 0.19 $&$0.92(2)$ \\
\hline
$(0,\frac{3}{2})$&$\Xi_{cc} \pi \to \Xi_{cc} \pi$ &$-0.140(14)$&$-0.143(24)$&$-0.16 \pm 0.02 $& $-0.10(2)$ \\
\hline \hline
\end{tabular}
\end{center}
\end{table}

\section{Summary}
We present a lattice QCD determination of the S-wave scattering lengths for the interactions between DCBs and GBs, by using $N_f=2+1$ CLQCD ensembles at pion masses of $M_\pi \simeq 210$ MeV and 300 MeV. The Lüscher's finite-volume method and the ERE formula are employed to extract the scattering parameters. It is found that our results of the scattering lengths are robust against discretization effects from heavy quarks. In addition, they are in good agreement with the available BChPT predictions, obtained at physical pion mass, in literature. Those BChPT predictions are determined with the values of LECs estimated by utilizing the HDA symmetry, indicating the goodness of the HDA symmetry between the DCB-$\phi$ and $D\phi$ systems. Our results provide essential inputs for future studies of the spectrum of double-heavy baryons.

\section{Acknowledgments}
This work is supported by Science Research Project of Hebei Education Department under Contract No.~QN2025063; by the Science Foundation of Hebei Normal University with Contract No.~L2025B09; by National Nature Science Foundations of China (NSFC) under Contract No.~12275076, No.~11905258, No.~12335002, No.~12175279, No.~12293060, No.~12293061; by the Science Fund for Distinguished Young Scholars of Hunan Province under Grant No.~2024JJ2007; by the Fundamental Research Funds for the Central Universities under Contract No.~531118010379.

\end{document}